\documentclass[conference]{IEEEtran}
\IEEEoverridecommandlockouts
\usepackage{cite}
\usepackage{amsmath,amssymb,amsfonts}
\usepackage{algorithmic}
\usepackage{graphicx}
\usepackage{textcomp}
\usepackage{xcolor}
\usepackage{bm}

\usepackage{amsmath,amssymb}
\usepackage{subfigure}

\usepackage{graphicx,graphics,color,psfrag}
\usepackage{cite,balance}
\usepackage{caption}
\allowdisplaybreaks
\usepackage{algorithm}
\usepackage{algorithmic}
\usepackage{accents}
\usepackage{amsthm}

\usepackage{url}
\usepackage[english]{babel}
\usepackage{multirow}
\usepackage{enumerate}
\usepackage{cases}
\usepackage{stfloats}
\usepackage{dsfont}
\usepackage{color,soul}
\usepackage{amsfonts}
\usepackage{fancyhdr}
\usepackage{hhline}
\usepackage{array,color}
\usepackage{mathtools}
\usepackage{pifont}

\newtheorem{remark}{Remark}

\def\BibTeX{{\rm B\kern-.05em{\sc i\kern-.025em b}\kern-.08em
    T\kern-.1667em\lower.7ex\hbox{E}\kern-.125emX}}
\begin{document}

\title{Channel Estimation for XL-IRS Assisted  Wireless Systems with Double-sided Visibility Regions}

\author{\IEEEauthorblockN{Chao Zhou\IEEEauthorrefmark{1}, Changsheng You\IEEEauthorrefmark{1},  Shiqi Gong\IEEEauthorrefmark{2}, Bin Lyu\IEEEauthorrefmark{3}, Beixiong Zheng\IEEEauthorrefmark{4}, and Yi Gong\IEEEauthorrefmark{1}} 
	\IEEEauthorblockA{\IEEEauthorrefmark{1}\text{Department of Electronic and Electrical Engineering}, \text{Southern University of Science and Technology}, Shenzhen, China,\\
	\IEEEauthorrefmark{2}\text{School of Cyberspace Science and Technology}, \text{Beijing Institute of Technology}, Beijing, China,\\
		\IEEEauthorrefmark{3}\text{School of Communications and Information Engineering}, \text{Nanjing University of Posts and Telecommunication}, Nanjing, China,\\
		\IEEEauthorrefmark{4}\text{School of Microelectronics}, \text{South China University of Technology}, Guangzhou, China,\\
	}
		Emails: zoe961992059@163.com, \{youcs, gongy\}@sustech.edu.cn, gsqyx@163.com, blyu@njupt.edu.cn, bxzheng@scut.edu.cn.
	}

\maketitle

\begin{abstract}
	In this paper, we study efficient channel estimation design for an extremely large-scale intelligent reflecting surface (XL-IRS) assisted multi-user communication systems, where both the base station (BS) and users are located in the near-field region of the XL-IRS. Two unique channel characteristics of XL-IRS are considered, namely, the \emph{near-field spherical wavefronts} and \emph{double-sided visibility regions} (VRs) at the BS and users, which render the channel estimation for XL-IRS highly challenging. To address this issue, we propose in this paper an efficient three-step XL-IRS channel estimation method.  Specifically, in the first step, an anchor node is delicately deployed near the XL-IRS to estimate the cascaded BS-IRS-anchor channel. Then, an efficient VR detection method is devised to estimate the VR information between the BS and XL-IRS. In this way, only the channels from the visible XL-IRS elements to the BS are estimated, thereby reducing the dimension of the cascaded BS-IRS-users channels to be estimated. 
	Third, by leveraging the common BS-IRS channel, the cascaded channels for all users are consecutively estimated accounting for the VRs of the IRS-user channels.
	Finally, numerical results are provided to demonstrate the effectiveness of our proposed channel estimation scheme as compared to various benchmark schemes.
\end{abstract}


\section{Introduction}

Recently, \emph{intelligent reflecting surface} (IRS) has emerged as a promising technology to smartly reconfigure the radio  propagation environment by dynamically controlling the signal reflections via its low-cost reflecting elements~\cite{RIS_Tutorial,you2024next,shao2022target}. To compensate for the product-distance path-loss in high frequency bands, one efficient approach is packing more and more reflecting elements, leading to the so-called extremely large-scale IRS (XL-IRS) \cite{you2024next}. Compared with the conventional IRS with a modest number of reflecting elements, XL-IRS introduces a fundamental change in the wireless channel modelling, shifting from the conventional far-field planar wavefronts to the new near-field spherical ones~\cite{NF_SphericalWave,Yunpu2022}.

In order to fully unleash the potential of XL-IRS in enhancing wireless communication performance, accurate channel state information (CSI) is indispensable yet challenging to acquire. It is worth that the existing works on IRS channel estimation (e.g.,~\cite{RIS_ChannelEst_LiuLiang,RIS_ChannelEst_TwoStage,RIS_ChannelEst_YuweiCS,RIS_ChannelEst_MVU,YouJSAC}) have been mostly based on the assumption of far-field channel modelling with spatial stationarity.  This, however, may not achieve desirable estimation performance in  XL-IRS systems, due to the channel mismatch (i.e., planer versus spherical wavefronts, and spatial stationary versus non-stationary). To address this issue, recent research efforts have been made to study efficient near-field channel estimation methods for XL-IRS systems. Specifically, for narrow-band systems, the authors in~\cite{XL-RIS_Est1} proposed to exploit the polar-domain sparsity to estimate the cascaded channels by developing a denoising convolutional neural network. This method was further extended in~\cite{XL-RIS_Est2} to estimate the XL-IRS channels in wide-band systems by leveraging the polar-domain sparsity to recover the cascaded channel.
Besides the polar-domain codebook based XL-IRS channel estimation methods~\cite{XL-RIS_Est1,XL-RIS_Est2}, a parameter estimation-based channel estimation scheme was proposed in~\cite{XL-RIS_Est3}, where a two-phase multi-user channel estimation scheme was devised  to recover the parameters of XL-IRS  hybrid-field channel. 
In addition to the spherical wavefronts, the visibility region (VR)-based channel estimation has also been studied for XL-IRS systems, where different IRS reflecting elements may encounter different propagation environments~\cite{NF_SphericalWave} (e.g., shadowing and blockages), resulting in the spatial non-stationarity.  For example, the authors in~\cite{XL-RIS_Est4-VR} proposed an efficient channel estimation method for XL-IRS assisted mmWave systems with VRs, while it assumed the simplified uniform linear array (ULA) IRS and only considered the user-sided VR. In~\cite{XL-RIS_Est5-VR}, both the VRs at the transceivers  were considered, where a U-shaped network based on the multilayer perceptron (MLP) architecture was employed to achieve  spatial non-stationary channel reconstruction.  However, the pilot overhead in~\cite{XL-RIS_Est5-VR} significantly increases when there is a large number of users.  

In view of the above works~\cite{RIS_ChannelEst_LiuLiang,RIS_ChannelEst_TwoStage,RIS_ChannelEst_YuweiCS,RIS_ChannelEst_MVU,YouJSAC,XL-RIS_Est1,XL-RIS_Est2,XL-RIS_Est3,XL-RIS_Est4-VR,XL-RIS_Est5-VR}, these still exist two key issues in XL-IRS channel estimation, when the XL-IRS has an extremely large aperture. First, different from the existing works that only considered the near-field IRS-user channel, both the BS and users may be located in the near-field region, thus rendering the cascaded channel modelling and channel estimation more complicated. Second, the spatial non-stationarity may concurrently exists in both the BS-IRS and IRS-user channels, termed as \emph{double-sided} VRs, where only a portion of XL-IRS elements are visible to the BS and users. To address these issues, in this paper, we study efficient channel estimation design for XL-IRS assisted multi-user systems, where both the BS and users are located in the near-field region of the XL-IRS. A three-step XL-IRS channel estimation scheme is proposed to effectively estimate the BS-IRS-user cascaded channels accounting for the non-uniform spherical wavefronts and double-sided VRs for the XL-IRS. Specifically, in the first step, an anchor node is delicately deployed to estimate the cascaded BS-IRS-anchor channel for decoupling the double-sided VRs. Then, the BS-sided VR information is accurately detected by an efficient VR detection method. As such, only the channels from the visible XL-IRS elements to the BS are estimated, thereby reducing the pilot overhead for estimating the cascaded BS-IRS-user channels. Finally, the common BS-IRS channel and BS-sided VR are utilized to consecutively estimate the cascaded BS-IRS-user channels with different user-sided VRs.
Numerical results showcase the superior performance of our proposed XL-IRS channel estimation method as compared to various benchmark schemes.


\section{System Model}

\begin{figure}
	\centering
	\includegraphics[width=0.7\linewidth]{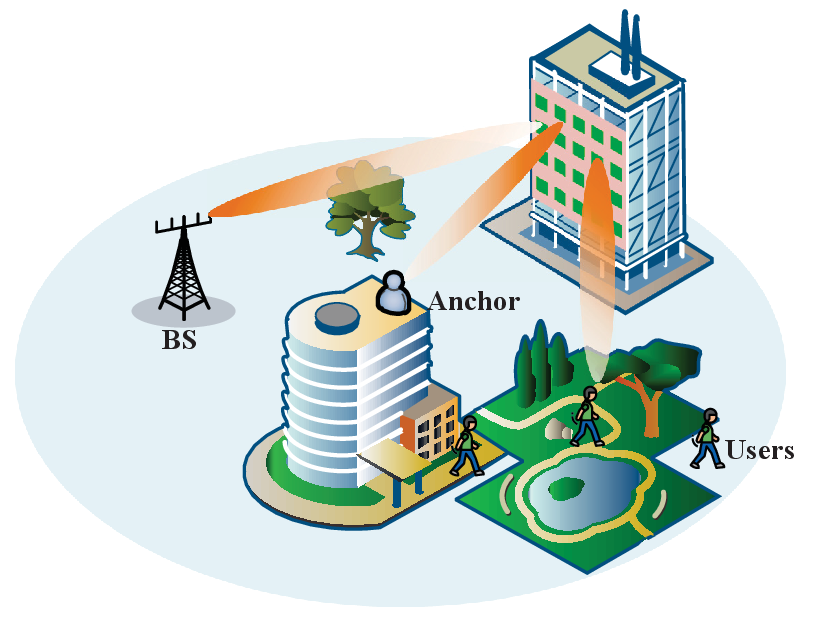}
	\caption{An XL-IRS assisted multi-user  system.}
	\label{System_model}
\end{figure}

We consider an XL-IRS assisted multi-user communication system as shown in  Fig.~\ref{System_model}, where 
an XL-IRS with $N =N_{x} \times N_{y}$ reflecting elements is deployed to assist the communications between an $M$-antenna BS and $K$ single-antenna users. Moreover, a single-antenna anchor node is placed near the XL-IRS to facilitate the cascaded XL-IRS channel estimation.

\subsection{Channel Model}
As shown in~\cite{NF_SphericalWave}, the Rayleigh distance of the XL-IRS system is given by $Z_{\rm B/U} = \frac{2\left(D_{\rm R}+D_{\rm B/U}\right)^2 }{\lambda}$, where $D_{\rm R}$ and $ D_{\rm B/U}$ denote the apertures of the XL-IRS and BS or users, respectively. As such, when $D_{\rm R}$ and $D_{\rm B/U}$ are sufficiently large, the users and BS are very likely to be located in the near-field region of the XL-IRS. Thus, we consider the general non-uniform spherical wave propagation model for the channels associated with the XL-IRS~\cite{NF_SphericalWave}. Specifically, the channel from the $n$-th XL-IRS reflecting element to the $m$-th BS antenna, denoted as $\check{\bm G}$, can be modeled as 
\begin{align}\label{multipath}
	&\check{\bm G}\left(m, n\right) = {\bm G}_{\text{LoS}}\left(m, n\right) +{\bm G}_{\text{NLoS}}\left(m, n\right) \nonumber \\
	&~~~~ = \beta_{m,n} e^{-j\frac{2\pi}{\lambda} d_{m,n}} +\sum_{\ell=1}^{L} \beta_{m,n}^{\left( \ell\right) } e^{-j\frac{2\pi}{\lambda}\left(  d_{m,\ell}+d_{n,\ell}\right) }, 
\end{align}
where $\beta_{m,n}$ denotes the complex-valued path gain, $d_{m,n}$ represents the corresponding transmission distance, and $\lambda$ is the carrier wavelength. Additionally, $\beta_{m,n}^{\left(\ell \right) }$ signifies the complex-valued path gain of the $\ell$-th path, while $d_{m,\ell}$ ($d_{n,\ell}$) denotes the distances between the $\ell$-th scatterer and the $m$-th antenna (between the $\ell$-th scatterer and the $n$-th reflecting element).
We assume that there exists a line-of-sight (LoS) dominant path due to the severe path-loss and shadowing in high-frequency bands.\footnote{The LoS path is adopted in this work due to the severe propagation attenuation of high frequency carriers~\cite{NF_SphericalWave}. However, the proposed estimation scheme can be extended to the multi-path scenario.} As such, the multi-path channel in~\eqref{multipath} can be approximated as $\check{\bm G} \approx {\bm G}_{\text{LoS}}$.
Due to the extremely large array aperture, only part of the XL-IRS reflecting elements are visible to some of the BS antennas due to the random obstructions such as trees or buildings. Accounting for such channel spatial non-stationarity, the channel from the XL-IRS to the BS can be rewritten as
\begin{align}
	\bm G = \check{\bm G}  \odot \bm {V}_{G}\in \mathbb{C}^{M\times N},
\end{align}
where $ \bm {V}_{G} \in \mathbb{C}^{M \times N}$ is the mask matrix with each of its elements being either zero or one. Specifically, $ \bm {V}_{G}\left(m,n\right)=0 $ indicates that the LoS path from the $n$-th XL-IRS element to the $m$-th BS antenna is obstructed (i.e., the $n$-th XL-IRS element is invisible to the $m$-th antenna).\footnote{Specifically, $\bm {V}_{G}\left(: ,n\right)=\bm 0 $ signifies that the $n$-th reflecting element is invisible to the entire BS, while $\bm {V}_{G}\left(m,:\right)=\bm 0$ indicates that the entire XL-IRS is invisible to the $m$-th antenna at the BS.} 
Similarly, we assume that only a portion of XL-IRS antennas are visible to the $k$-th user. Then the user-IRS channel can be modeled as 
\begin{align}
	\bm {r}_{k} = \bm {\beta}_{k} \odot \bm {\alpha}_{k} \odot \bm{v}_{k}\in \mathbb{C}^{N \times 1},
\end{align}
where $ \bm {\beta}_{k}=\left[\beta_{k,1}, \ldots, \beta_{k,n}, \ldots, \beta_{k,N}\right]^T\in \mathbb{C}^{N\times 1}   $ represents the complex path gain vector from the $k$-th user to the XL-IRS, $ \bm {\alpha}_{k} =\left[\alpha_{k,1}, \ldots, \alpha_{k,n}, \ldots, \alpha_{k,N}\right]^T\in \mathbb{C}^{N\times 1}$  with $ \alpha_{k,n} = e^{-j2\pi\frac{d_{n,k}}{\lambda}}$,  $d_{n,k}$ is the distance from the $n$-th XL-IRS reflecting element to the $k$-th user, and $ \bm{v}_{k} \in\mathbb{C}^{M\times1}$ is the VR indicator for the channel with each of its elements being either zero or one. Accordingly, the double-sided VR-based cascaded channel from the $k$-th user to the BS via the XL-IRS can be expressed as
\begin{align}
	\bm H_{k} = \bm G \text{diag}\left(\bm {r}_{k} \right) \in \mathbb{C}^{M\times N}.
\end{align}

Since $\bm H_{k}$ contains the VR information of both the BS and $k$-th user (i.e., $\bm {V}_{G} $ and $\bm{v}_{k}$), the estimation of cascaded channels becomes more intricate. 
To accurately estimate the VR-based cascaded channels with low pilot overhead, we propose to deploy a single-antenna anchor node to assist in decoupling the VR information of the BS and users in the cascaded channel. Specifically, the anchor is elaborately positioned to prevent obstruction between the XL-IRS and the anchor.\footnote{The anchor node is elaborately placed in accordance with the position of the XL-IRS, such as the roof of a tall building, to ensure the unobstructed LoS path between the XL-IRS and anchor.} Accordingly, the reflected channel from the anchor to the XL-IRS can be expressed as $\bm {r}_{\rm a} =\bm {\beta}_{\rm a} \odot \bm {\alpha}_{\rm a} \in \mathbb{C}^{N \times 1}$,
where $ \bm {\beta}_{\rm a}=\left[\beta_{{\rm a},1}, \ldots, \beta_{{\rm a},n}, \ldots, \beta_{{\rm a},N}\right]^T\in \mathbb{C}^{N\times 1}   $ represents the complex path gain vector from the anchor to the XL-IRS, $ \bm {\alpha}_{\rm a} =\left[\alpha_{{\rm a},1}, \ldots, \alpha_{{\rm a},n}, \ldots, \alpha_{{\rm a},N}\right]^T\in \mathbb{C}^{N\times 1}$  with $ \alpha_{{\rm a},n} =e^{-j2\pi\frac{d_{\rm a}}{\lambda}}$, and $d_{\rm a}$ is the distance from the $n$-th XL-IRS reflecting element to the anchor. Accordingly, the cascaded channel from the anchor to the BS via the XL-IRS can be expressed as 
\begin{align}
	\bm {H}_{\rm a} = \bm G \text{diag}\left(\bm {r}_{\rm a} \right) \in \mathbb{C}^{M\times N},
\end{align}
which only contains the VR information of the BS (i.e., $ \bm V_{G} $).

\subsection{Transmission Protocol}
A practical transmission protocol is considered, in which each channel coherence time $T$ is divided into two phases. During the channel estimation phase, the users and anchor node send $T_{\rm e}$ pilot symbols to facilitate channel estimation. The second phase is utilized for data transmissions by jointly designing the BS active beamforming and  XL-IRS passive beamforming. Let $\bm w_{k}$ and $\bm {\phi}$ denote the transmit beamforming vector and reflection vector, respectively. As such, the effective sum-rate in bits/second/Hertz (bit/s/Hz) is given by
\begin{align}
	R_{\rm eff} = \left(1- \frac{T_{\rm e}}{T} \right) \sum_{k=1}^{K}\log_2\left(1+ \text{SINR}_{k} \right), 
\end{align}
where $ \text{SINR}_{k} = \frac{|\bm {\phi}^H \bm H_k^H \bm {w}_{k}|^2}{\sum_{k^{'}\neq k}^{K} |\bm{\phi}^H \bm H_{k}^H \bm{w}_{k^{'}} |^2+\sigma_{k}^2}$ and $\sigma_{k}^2$ denotes the noise power at the $k$-th user.

\section{Proposed XL-IRS Estimation Method \\ under Double-sided VRs}\label{section3}

In this section, we propose an efficient anchor-assisted channel estimation method for the XL-IRS assisted multi-user  systems with double-sided VRs. 

Note that for the considered XL-IRS systems with double-sided VRs, the existing IRS channel estimation methods such as~\cite{RIS_ChannelEst_LiuLiang,RIS_ChannelEst_TwoStage,RIS_ChannelEst_YuweiCS,RIS_ChannelEst_MVU} cannot be directly applied due to the following reasons. First, existing multi-user IRS channel estimation methods by exploiting the common BS-IRS channel~\cite{RIS_ChannelEst_LiuLiang} cannot be directly applied, since it may result in severe channel estimation errors.  This is attributed to the fact that the proportional relationship between the cascaded channels does not always hold because of different VRs for different users. Second, the full-duplex BS based channel estimation method~\cite{RIS_ChannelEst_TwoStage} may incur severe estimation errors for the common BS-IRS channel estimation, since some pilot symbols transmitted by the BS may not be received due to the VR at the BS. Third, the double-sided VRs for the XL-IRS destroys the sparsity of cascaded channels, rendering the compressed sensing (CS)-based channel estimation methods (e.g., \cite{RIS_ChannelEst_YuweiCS}) being possibly ineffective. Fourth, the pilot overhead in~\cite{RIS_ChannelEst_MVU} becomes unaffordable  as the number of users increases due to the extremely large number of XL-IRS reflecting elements.  

To address theses issue, we propose to deploy an anchor node for decoupling the double-sided VRs and estimate its cascaded channel. Based on the estimated cascaded channel for the anchor, the VR information of the BS is detected, which is then utilized to reduce the dimension of cascaded channels that need to estimate.


\begin{figure}
	\centering
	\includegraphics[width=1.1\linewidth]{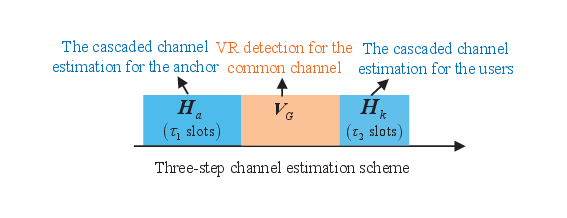}
	\caption{Three-step XL-IRS channel estimation scheme.}
	\label{frame}
\end{figure} 
\subsection[${H}_{\rm a}$]{Estimation of Anchor Channel $\bm{H}_{\rm a}$}
In the first step, the anchor  node consecutively sends pilots to the BS while the XL-IRS dynamically adjusts its reflecting elements. Accordingly, the received signal at the  $i$-th time slot, denoted as $\bm y_{i}\in \mathbb{C}^{M \times 1}$, is given by
\begin{align}
	\bm {y}_{i} =\sqrt{p} \bm H_{\rm a} \bm {\phi}_{i}  x_{i} + \bm z_{i},~1 \le i \le \tau_1,
\end{align}
where $p$ denotes the transmit power of the anchor, $\bm {\phi}_{i} =\left[\phi_{1,i}, \ldots, \phi_{n,i}, \ldots, \phi_{N,i} \right]^T \in \mathbb{C}^{N \times 1}$ represents the XL-IRS reflection vector at the $i$-th time slot with $\left| \phi_{n,i}\right|=1 $, $x_{i}$ is the pilot symbol transmitted by the anchor, and $\bm z_{i}$ denotes the additive white Gaussian noise (AWGN) at the BS during the $i$-th time slot with $\bm z_{i} \sim \mathcal{CN}\left(\bm 0, \sigma^2 \bm I_{M}\right) $.  It can be easily shown that to estimate the cascaded anchor-BS channel, it at least requires $\tau_{1} = N$ pilot symbols at the anchor. To minimize the training overhead, we set $\tau_{1}=N$.  Accordingly, by setting $x_i=1$ without loss of generality, the overall received signal at the BS in the first step can be expressed as
\begin{align}\label{Re_pilot_1}
	\bm Y^{\left(1 \right)} = \left[\bm y_{1},\ldots, \bm y_{N} \right] = \sqrt{p} \bm H_{\rm a} \bm {\Phi}_{1} + \bm Z_{1},
\end{align}
where $\bm \Phi_{1} =\left[\bm {\phi}_{1},\ldots, \bm {\phi}_{N} \right]\in \mathbb{C}^{N \times N} $ and $\bm Z_{1} =\left[\bm {z}_{1},\ldots, \bm {z}_{N} \right] \in \mathbb{C}^{N \times N} $.  According to~\cite{BeixiongCE,YouJSAC}, $\bm \Phi_{1}$ should be configured as an $N$-dimensional discrete Fourier transform (DFT) matrix to minimize the channel estimation error. Thus, we have $\bm \Phi_{1} \left( \bm \Phi_{1}\right)^H =N \bm I_N $. Based on the LS estimation scheme, the estimated cascaded channel for the anchor can be obtained as 
\begin{align}\label{Es_Anchor}
	\hat{\bm {H}}_{\rm a} = \frac{1}{\sqrt{p}N} \bm Y^{\left(1 \right) }  \left( \bm \Phi_{1}\right)^H.
\end{align}

\subsection[${V}_{G}$]{VR Detection of $\bm{V}_{G}$}
Since the entire XL-IRS is visible to the anchor, the estimated cascaded channel for the anchor contains the full VR information of the BS. Therefore,  after obtaining $\hat{\bm {H}}_{\rm a}$, we can recover the mask matrix $\bm V_{G}$ via efficient VR detection. Specifically, by substituting~\eqref{Re_pilot_1} into~\eqref{Es_Anchor}, the estimated cascaded channel for the anchor can be rewritten as
\begin{align}
	\hat{\bm {H}}_{\rm a} = {\bm {H}}_{\rm a} + \Delta \bm {H}_{\rm a},
\end{align}
where $ \Delta \bm {H}_{\rm a} = \frac{\bm Z_{1}\left(\bm {\Phi_{1}} \right)^H}{\sqrt{p}N} $ represents the channel estimation error, with each element distributed as $ \Delta \bm {H}_{\rm a}\left(m,n\right) \sim \mathcal{CN}\left(0, \frac{\sigma^2}{p N} \right) $. Note that  when $\bm V_{G}\left(m,n\right) =0 $, we have $\bm {H}_{\rm a}\left(m,n\right) =0$
and  $\hat {\bm H}_{\rm a}\left( m, n\right)= \Delta \bm {H}_{\rm a}\left(m,n \right) $.
Based on the distribution of $ \Delta \bm {H}_{\rm a}\left(m,n\right)$, we estimate the mask matrix $\bm V_{G}$ for the common channel as follows by using the threshold based on energy detection.
\begin{equation}
	\label{VR_Detection}
	\hat{\bm V}_{G}\left(m,n \right)=\left\{ \begin{aligned}
		& 1,~ \left| \hat{\bm {H}}_{\rm a}\left(m,n\right)\right|^2 > \zeta; \\
		& 0, ~ \left| \hat{\bm {H}}_{\rm a}\left(m,n\right)\right|^2 \le \zeta,\\
	\end{aligned} \right.
\end{equation}
where $\hat{\bm V}_{G}$ represents the estimated VR mask matrix of the BS, $\zeta = \frac{3\sigma^2}{p N}$ is the threshold for VR detection. If all elements of the $n$-th column of $\hat{\bm V}_{G}$ are zero, it implies that the $n$-th XL-IRS reflecting element is invisible to all the BS antennas. As such, the pilot signal transmitted by the anchor or users through the $n$-th XL-IRS reflecting element cannot be received at the BS. This is reasonable since the BS has relatively smaller number of antennas and all antennas are densely packed with half-wavelength spacing and hence may share the same/similar VR information with individual IRS reflecting elements.
 In other words, the cascaded channel from the users to the BS, reflected by the $n$-th XL-IRS reflecting element, does not need to be estimated, which provides an opportunity for reducing the pilot overhead when estimating the cascaded channels for all the $K$ users.

Let $\tilde{N} ( \tilde{N} \le N) $ denote the total number of visible XL-IRS reflecting elements to the BS and the  corresponding  index set is defined as $\Omega_G = \{{\chi}_{1},\ldots,{\chi}_{\tilde{n}},\ldots,{\chi}_{\tilde{N}} \}$, where ${\chi}_{\tilde{n}}$ represents the index of the $\tilde{n}$-th visible XL-IRS reflecting element to the BS with $1 \le \tilde{n} \le \tilde{N} $. Then, we remove the columns in $\bm H_{\rm a}$ and $\bm H_{k}$ that do not require channel estimation (i.e., $n \notin \Omega_G $), and the reduced-dimensional cascaded channels can be respectively written as  $\tilde{\bm H}_{\rm a}  = \tilde{\bm G}  \text{diag}\left(\tilde{\bm r}_{\rm a} \right)  \in \mathbb{C}^{M\times\tilde{N}}$ and $\tilde{\bm H}_{k} = \tilde{\bm G}  \text{diag}\left(\tilde{\bm r}_{k} \right) \in \mathbb{C}^{M\times\tilde{N}}$, where $ \tilde{\bm G} \left(:,\tilde{n}\right) = \bm G \left(:,\chi_{\tilde{n}} \right)   $, $\tilde{\bm r}_{\rm a}\left(\tilde{n} \right)  = \bm r_{\rm a}\left(\chi_{\tilde{n}} \right)  $, and $\tilde{\bm r}_{k}\left(\tilde{n} \right)  = \bm r_{k}\left(\chi_{\tilde{n}} \right)  $. Thus, we have 
\begin{align}
	&\tilde{\bm H}_{\rm a} \left( :,\tilde{n}\right)  = \bm {H}_{\rm a}\left(:,\chi_{\tilde{n}}\right), \\
	 & \tilde{\bm H}_{k} \left( :,\tilde{n}\right)  = \bm {H}_{k}\left(:,\chi_{\tilde{n}}\right).
\end{align}
\begin{remark}[Exploiting VRs for reducing pilot overhead] 
	\rm Note that for the full cascaded channels of the users $\bm {H}_{k}$, the total number of estimated channels required is $KMN$, \emph{while those in the reduced-dimensional cascaded channels $ \tilde{\bm H}_{k} $ are $KM \tilde{N}$} only, due to the presence of  VR of the common channel. This implies that a \emph{lower pilot overhead} is required to estimate the cascaded channels for the users in the third step, thereby avoiding unnecessary pilot transmissions. In other words, as some reflecting elements of the XL-IRS are not visible to the entire BS, there is no need to estimate the corresponding links from all users to these elements, which provides an opportunity for further reducing the pilot overhead.
\end{remark}

\subsection[${H}_{k}$]{Estimation of User Channels $\bm{H}_{k}$}
After the VR detection, the common BS-IRS channel  is leveraged to estimate the cascaded channels for all the users. Specifically,  given $\tilde{\bm r}_{\rm a}$, the reduced-dimensional channels for the $k$-th user is estimated via~\cite{RIS_ChannelEst_LiuLiang}
\begin{align}
	\tilde{\bm r}_{k} = \tilde{\bm \lambda}_k  \odot \tilde{\bm r}_{\rm a},
\end{align}
where $ \tilde{\bm \lambda}_{k} =\left[\tilde{\lambda}_{k,1},\ldots,\tilde{\lambda}_{k,\tilde{n}},\ldots,\tilde{\lambda}_{k,\tilde{N}}   \right]^T  $ is the corresponding scaling vector with $ \tilde{\lambda}_{k,\tilde{n}}= \frac{\tilde{\bm r}_{k}\left(\tilde{n} \right) }{\tilde{\bm r}_{\rm a} \left(\tilde{n} \right)  } $. 
Thus, the received signal at the BS transmitted by each user at the $i$-th time slot can be expressed as
\begin{align}\label{slot_signal-2}
	\bm y_{i}& =\sum_{k=1}^{K} \sqrt{p_{\rm u}} \tilde{\bm H}_{k} {\tilde{\bm  \phi}}_{i}  x_{k,i} + \bm z_{i} \nonumber \\
	& = \sum_{k=1}^{K} \sqrt{p_{\rm u}} \tilde{\bm H}_{\rm a}\text{diag}\left( \tilde{\bm \lambda}_k \right)  {\tilde{\bm  \phi}}_{i}  x_{k,i} + \bm z_{i}\nonumber \\
	&= \sum_{k=1}^{K} \sqrt{p_{\rm u}} \tilde{\bm H}_{\rm a} \text{diag}\left( {\tilde{\bm  \phi}}_{i}\right) \tilde{\bm \lambda}_k   x_{k,i} + \bm z_{i} \nonumber \\
	& = \sqrt{p_{\rm u}} \bm {\Xi}_{i} \tilde{\bm \lambda}+ \bm z_{i},~\tau_1 < i \le \tau_1+\tau_2,
\end{align}
where $\bm {\Xi}_{i} =\left[  x_{1,i}\tilde{\bm H}_{\rm a} \text{diag}\left( {\tilde{\bm  \phi}}_{i}\right),\ldots, x_{K,i}\tilde{\bm H}_{\rm a} \text{diag}\left( {\tilde{\bm  \phi}}_{i}\right)  \right]\in \mathbb{C}^{M \times \tilde{N}K} $,  $\tilde{\bm \lambda} = \left[\tilde{\bm \lambda}_1;\ldots;\tilde{\bm \lambda}_K \right] \in \mathbb{C}^{\tilde{N}K \times 1}$, and $p_{\rm u} $ is the transmit power at the users.
It can be shown  that the unknown variables (i.e., $ \tilde{\bm \lambda}_k $) in~\eqref{slot_signal-2} are $K \tilde{N}$ only, which depends on the number of users and the visible XL-IRS reflecting elements to the BS. 
Hence, the pilot overhead in the third step can be significantly reduced for two  reasons. Firstly, the correlation between $\tilde{\bm r}_{\rm a}$ and $\tilde{\bm r}_{k}$ is utilized to estimate $\tilde{\bm H}_{k}$ as they share a common channel $\tilde{\bm G}$ with $\tilde{\bm H}_{\rm a}$. Secondly, the existence of VRs in the common channel reduces the number of channel parameters to be estimated. During $\tau_2$ time slots of the third step, the  signals received by the BS can be concatenate as  
\begin{align}\label{thirdStep}
	\bm y^{\left(3 \right) } &= \left[\bm y_{\tau_1+1};\ldots;\bm y_{\tau_1+i};\ldots;\bm y_{\tau_1+\tau_2}  \right] =\sqrt{p_k} \bm \Xi \tilde{\bm \lambda} + \bm z  \nonumber \\
	& =\sqrt{p_k}\left[\bm {\Xi}_{\tau_1+1};\ldots;\bm {\Xi}_{\tau_1+i};\ldots; \bm {\Xi}_{\tau_1+\tau_2}\right]  \tilde{\bm \lambda} + \bm z,
\end{align}
where $\bm z =\left[\bm {z}_{\tau_{1}+1};\ldots;\bm {z}_{\tau_{1}+i};\ldots;\bm {z}_{\tau_{1}+\tau_2}  \right]  $ and $\tau_1 < i \le \tau_1+\tau_2$.  Then it can be shown that  $\tau_2 M \ge K \tilde{N}$ pilot symbols are required to estimate the cascaded user-BS channel parameters by constructing a full-rank matrix $\bm \Xi$. 
According to~\cite{RIS_ChannelEst_LiuLiang}, by carefully designing the pilot symbol and reflection coefficient, we can obtain the full-rank matrix $\bm \Xi$ with $\tau_2 =\lceil\frac{K \tilde{N}}{M} \rceil$, and the estimated scaling vector can be obtained as 
\begin{align}
	\bar{\bm \lambda} = \frac{1}{\sqrt{p_{\rm u}}}\bm \Xi^{\dagger} \bm y^{\left(3 \right) }.
\end{align}
Based on the estimated scaling vector, the reduced-dimensional cascaded channels can be estimated as
\begin{align}
	\bar{\bm H}_{k} = \bar{\bm H}_{\rm a}\text{diag}\left( \bar{\bm \lambda}_k \right),
\end{align} 
where $ \bar{\bm H}_{\rm a} \left( :,\tilde{n}\right)  = \hat {\bm {H}}_{\rm a}\left(:,\chi_{\tilde{n}}\right) $. Based on $\bar{\bm H}_{k} $, the $\chi_{\tilde{n}}$-th row of the estimated cascaded channel for the $k$-th user, denoted as $\hat{\bm H}_{k}\left(:,\chi_{\tilde{n}}\right)$, can be obtained as $ \hat{\bm H}_{k}\left(:,\chi_{\tilde{n}}\right)= \bar{\bm H}_{k}\left(:,\tilde{n}\right)$, where  $ \chi_{\tilde{n}} \in  \Omega_G$ and the remained rows are set as $\bm 0_{M \times 1}$. 
\begin{remark}[Channel estimation overhead]
	\rm
	Based on the above three channel estimation steps, the total number of required pilot symbols can be obtained as $\frac{\tau_1}{\kappa} + \tau_2 = \frac{N}{\kappa}+ \lceil\frac{K \tilde{N}}{M} \rceil$ with $\kappa \gg 1$. When $\kappa  \to \infty$, the average pilot overhead can be approximated as $ \lceil\frac{K \tilde{N}}{M} \rceil$. While for the proposed scheme without VR detection, the average pilot overhead is $ \lceil\frac{KN}{M} \rceil$. Furthermore, compared with the common channel based~\cite{RIS_ChannelEst_LiuLiang} and the DFT based~\cite{RIS_ChannelEst_MVU} channel estimation methods that need $N + \lceil\frac{ ( K-1 )N}{M}\rceil$ and $KN$ pilot symbols, respectively, our proposed anchor-based near-field channel estimation method by exploiting the double-sided VRs can significantly reduce the channel estimation overhead.
\end{remark}

\begin{remark}[Effects of double-sided VRs]
	\rm
	Due to the existence of double-sided VRs for the XL-IRS, the estimation of cascaded channels is divided into two parts to decouple the double-sided VRs. Specifically, the VR for the BS destroys the sparsity property of the common channel between the XL-IRS and BS, hence posing a challenge for estimating the common channel with low pilot overhead. On the other hand, the VRs for the users break the proportional relationship between reflected channels (i.e., $\bm {r}_{k}$), rendering the estimation scheme in~\cite{RIS_ChannelEst_LiuLiang} ineffective. Fortunately, the VR detection by the proposed channel estimation method can accurately identify the VR information of the BS, which can then be utilized to reduce the dimension of the cascaded channel  to be estimated. 
	Note that the proportional relationship between $\bm {r}_{\rm a}$ and $\bm {r}_{k}$ is used to further reduce the pilot overhead, even in the presence of different VRs for different user.
\end{remark}

\begin{remark}[Channel estimation error]\label{error}
	\rm Since the estimated reduced-dimensional cascaded channels for users are correlated with $\bar{\bm H}_{\rm a}$, the accuracy of channel estimation for $\tilde{\bm H}_{\rm a}$ and VR detection will affect the channel estimation accuracy for $\bar{\bm H}_{k}$. Given that the estimation error for ${\bm H}_{\rm a}$ (i.e., $\Delta \bm {H}_{\rm a}$) is inversely proportional to $p$ and $N$, to reduce the estimation error for $\bar{\bm H}_{k}$, \emph{more transmit power} and \emph{pilot symbols} can be allocated at the anchor to enhance the accuracy of channel estimation for $ \hat{\bm H}_{\rm a} $ in the first step. 
\end{remark}

\section{Numerical Results}\label{section-IV}

In this section, numerical results are present to validate the effectiveness of our proposed XL-IRS channel estimation scheme. The simulation parameters are set as in Table~\ref{SimulationParameters}, and all the $K$ users are located within circles centered at $(0,20,0)$ with a radius of $ 100 $ meters. Moreover, three benchmarks are considered for performance comparison:
\begin{itemize}
	\item {\emph{Common channel based scheme}:} The multi-user channels are estimated by the channel estimation scheme proposed in~\cite{RIS_ChannelEst_LiuLiang}.  The cascaded channel for a typical user is firstly estimated. Subsequently, the cascaded channels for other users are estimated by leveraging the common channel between the XL-IRS and the BS for each user.
	
	\item {\emph{DFT based scheme}:} The cascaded channel for each user is estimated by the scheme in~\cite{RIS_ChannelEst_MVU}, which achieves the minimum channel estimation error through DFT based IRS reflection pattern.
	
	\item {\emph{Proposed scheme without (wo.) VR Detection}:} The cascaded channel for the anchor is first estimated, followed by the estimation of the cascaded channel for each user without the process of  VR detection.

\end{itemize}
\begin{table}[t!]
		\renewcommand\arraystretch{1}
		\centering
		\caption{Simulation Parameters}
		\label{SimulationParameters}
		\begin{tabular}{c|c}
			\hline 
			\textbf{Parameter} & \textbf{Value}          
			\\ \hline Carrier wavelength, $\lambda$ & $0.03$ m
			\\ \hline Number of antennas at the BS, $M$ & $128$ 
			\\ \hline Number of reflecting elements at the XL-IRS, $N$ & $480$ 
			\\ \hline Transmit power at the anchor, $p$  & $30 $  dBm
			\\ \hline Transmit power at the users, $p_{\rm u}$  & $30 $ dBm 
			\\ \hline Rayleigh distance, $Z_{\rm B}$ and $Z_{\rm U}$,  & $1098$  m and $856$ m
			\\ \hline Location of BS, & $ \left(100,0,20\right) $  m
			\\ \hline Location of XL-IRS, & $ \left(0,0,50\right) $  m
			\\ \hline Location of anchor, & $ \left(20,20,50\right) $  m
			\\ \hline
		\end{tabular}
\end{table}



\begin{figure}[t!]
	\centering
	\includegraphics[width=0.75\linewidth]{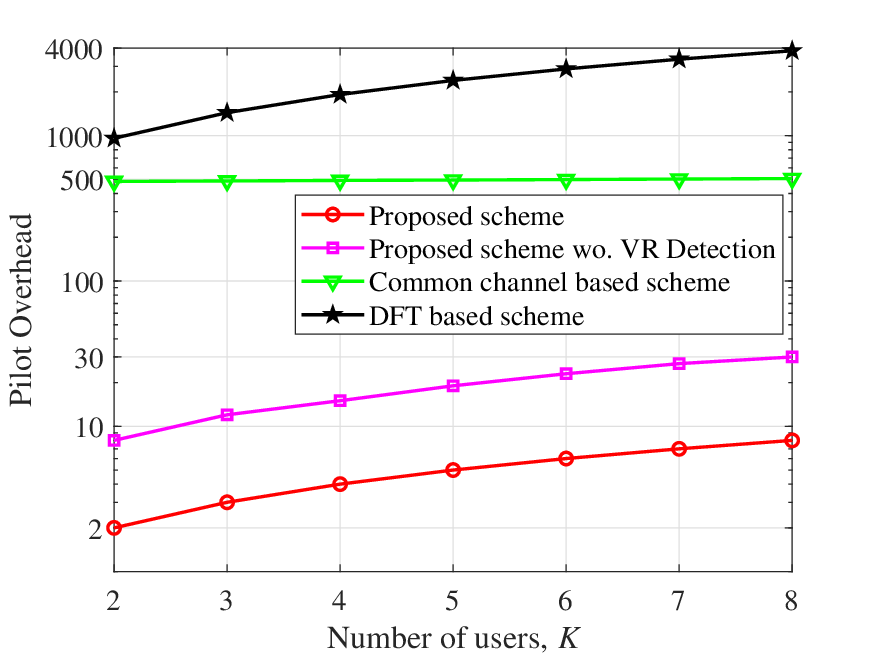}
	\caption{ Pilot overhead versus  number of users, $K$.}
	\label{Overhead}
\end{figure}

In Fig.~\ref{Overhead}, we plot the pilot overhead versus the number of users by different estimation schemes. Several main observations are made as follows. First, the pilot overhead for different schemes monotonically increases with the number of users. Second, by exploiting the common BS-IRS channel, our proposed three-step XL-IRS channel estimation scheme significantly reduces the channel estimation overhead of the DFT based scheme (e.g., $ 8 $ versus $ 3840 $ for $K=8$). Moreover, by further leveraging the VR information at the BS side, the proposed scheme achieves smaller pilot overhead than the scheme without VR detection (e.g., $8$ versus $30$ for $K=8$).


\begin{figure}[t!]
	\centering
	\includegraphics[width=0.75\linewidth]{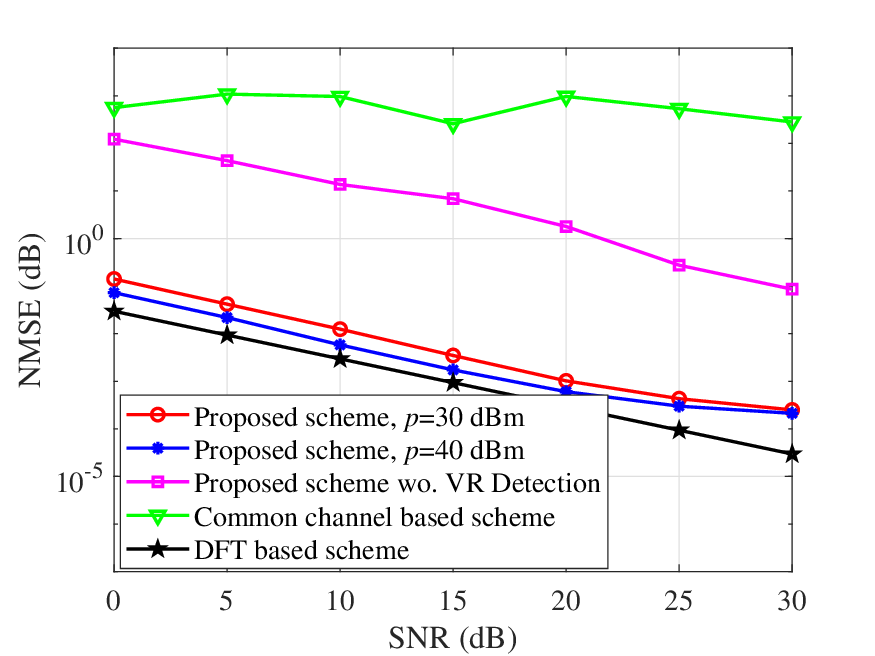}
	\caption{NMSE versus SNR.}
	\label{NMSE}
\end{figure}

Fig.~\ref{NMSE} illustrates the effects of SNR on the normalized mean square error (NMSE) of the cascaded channels for all schemes, which defined as $$\text{NMSE}=\frac{{\mathbb{E}\left\lbrace \sum_{k=1}^{K} \left| \left| \hat{\bm H}_{k} -{\bm H}_{k}   \right| \right|_F^2 \right\rbrace }}{{\mathbb{E}\left\lbrace \sum_{k=1}^{K} \left| \left|{\bm H}_{k}   \right| \right|_F^2 \right\rbrace}}. $$
It is evident that the DFT based scheme achieves the lowest NMSE as SNR increases, albeit at the expense of excessive pilot overhead (see Fig.~\ref{Overhead}). Note that our proposed scheme can achieve near-optimal estimation accuracy with the DFT-based scheme.
Moreover, the performance gap between the proposed scheme and the DFT based scheme can be reduced by increasing the transmit power at the anchor for improving the estimation accuracy of the cascaded channel in the first step, which aligns with~\textbf{Remark \ref{error}}.
Furthermore, compared to the scheme without VR detection, our proposed scheme significantly enhances channel estimation accuracy. This suggests that VR detection can be employed to  not only reduce the pilot overhead but also improve channel estimation accuracy.

\begin{figure}[t!]
	\centering
	\includegraphics[width=0.75\linewidth]{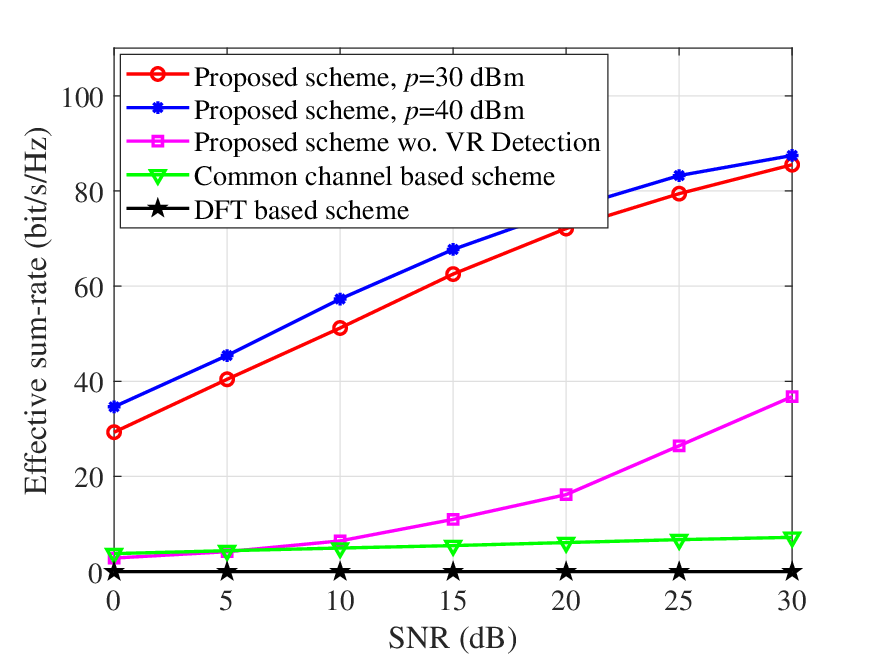}
	\caption{Effective sum-rate versus SNR.}
	\label{Eff-SR}
\end{figure}

In Fig.~\ref{Eff-SR}, we investigate the effective sum-rate versus the SNR across various estimation schemes. It is obvious that our proposed scheme attains the highest effective sum-rate, which showcases its advantage to achieve high data rates with minimal pilot overhead. In contrast, the DFT based scheme achieves the highest estimation accuracy at the cost of excessive pilot overhead, which results in less time for data transmission and achieve low effective sum-rate.

\section{Conclusions}


In this paper, we proposed an efficient three-step channel estimation scheme for XL-IRS assisted multi-user systems, by taking into account non-uniform spherical wavefronts and double-sided VRs at the BS and users. Specifically, by considering the double-sided VRs of both the BS and users in the cascaded channels, an anchor node was strategically deployed in the first step to facilitate accurate estimation of the cascaded BS-IRS-anchor channel. Leveraging the estimated channel for the anchor, the VR information of the BS was then detected and utilized to reduce the number of channel parameters that need to be estimated for each user. Finally, the proportional relationship between the cascaded channels for anchor and each user was exploited  to estimate the reduced-dimensional cascaded channels with different user-sided VRs. Numerical results validated that our proposed scheme can enhance the channel estimation accuracy with reduced pilot overhead.

\bibliographystyle{IEEEtran}
\bibliography{v1.bib}

\end{document}